\documentstyle[aps,eqsecnum,twocolumn,epsf,array,epsfig,psfig]{revtex}
\newcommand{\be}{\begin{equation}}
\newcommand{\ee}{\end{equation}}
\newcommand{\ben}{\begin{eqnarray}}
\newcommand{\een}{\end{eqnarray}}

\begin{document}
\setlength{\topmargin}{-2ex}

\title{Gauge Fluctuations in Superconducting Films}
\author{L. M. Abreu, A. P. C. Malbouisson and I. Roditi}
\address{{\it Centro Brasileiro de Pesquisas F\'{\i}sicas,
Rua Dr. Xavier Sigaud 150, 22290-180, Rio de Janeiro, RJ, Brazil}\\
{\it E-mail: lmabreu@cbpf.br, adolfo@lafex.cbpf.br,
roditi@cbpf.br}\\}

\maketitle

\begin{abstract}
\noindent In this paper we consider a superconducting film 
modeled by the Ginzburg-Landau model, confined between two
parallel planes a distance $L$ apart from one another. Our
approach is based on the Gaussian 
effective potential in the transverse unitarity gauge, which
allows to treat  gauge contributions in a compact form. Using
techniques from dimensional and $zeta$-function regularizations,
modified by the confinement conditions, we investigate the
critical temperature as a function of the film thickness $L$.
The contributions from the scalar self-interaction and from the gauge fluctuations
are clearly identified. The model suggests the existence of a minimal critical
thickness below which superconductivity is suppressed.\\
\\
\noindent PACS number(s): 74.20.-z, 05.10Cc, 11.25.Hf
\end{abstract}

\section{Introduction}

It is currently assumed to be a good approximation to neglect magnetic thermal 
fluctuations in the Ginzburg-Landau (GL) model, to explore general features of 
superconducting transitions. However, this approximation excludes the study 
of the so called 
charged phase transitions. They only can be investigated when fluctuations of 
the gauge field are taken into account, which make explicitly appear in the thermodynamic 
quantities the coupling constant for the interaction between the 
scalar field and the gauge field (the charge of the boson).
 This is a hard problem to be directly faced and 
many attempts have been done to go beyond the pioneering work of ref.\cite{halperin}.
In what concerns physical situations already present in the literature, a
large amount of work has been done on the Ginzburg-Landau model applied to
the study of superconductors, both in the single component and in the $N$ 
-component versions of the model, using the renormalization group approach.
 The interested reader can find an
account on the state of the subject for both type-I and type-II
superconductors and related topics in Refs. \cite{Lawrie,Lawrie1,Afleck,Brezin,Radz,Moore,Flavio1}.

In this paper, we intend to take into account gauge fluctuations
in a study of superconducting films. We consider the Ginzburg-Landau model,
the system being submitted to the constraint of confinement between two
parallel planes a distance $L$ apart from one another. From a physical point
of view, for dimension $d=3$ and introducing temperature by means of the mass term in
the Hamiltonian, this corresponds to a film-like material. We investigate the behaviour of the system taking into account gauge fluctuations, which means that charged transitions 
are included in our work. We are particularly interested in the problem of how the 
critical behaviour depends on the film thickness $L$. This study is done
by means of the Gaussian Effective Potential (GEP) as developed in Refs. \cite{Gep1,Gep2,Gep3,Gep4}, together with a spatial compactification mechanism
introduced in  recent publications\cite{Ademir,JMario}.

Effects of spatial boundaries on the behaviour of physical
systems appears in several forms in the literature. For instance, there are
systems that present defects, as domain walls created for instance in the
process of crystal growth by some prepared circumstances. At the level of
effective field theories, in many cases this can be modeled by considering a
Dirac fermionic field whose mass changes sign as it crosses the defect, what
means that the domain wall can be interpreted as a kind of a critical
boundary \cite{FAdolfo,Cesar}. Questions concerning stability and the
existence of phase transitions may also be raised if we enquire about the
behaviour of field theories as function of spatial boundaries. The existence
of phase transitions would be in this case also associated to some spatial
parameters describing the breaking of translational invariance, in our case
the distance $L$ between planes confining the system (a superconducting 
film of thickness $L$). In particular the
question of how the superconducting critical temperature could depend on the
thickness of the film can be raised.

In the next Section, we apply the functional approach of the  
Gaussian effective potential formalism to 
the Ginzburg-Landau model, obtaining the mass, which obeys a generalized "Gaussian" 
Dyson-Schwinger equation. In Section {\bf III}, we extend the formalism of Section
{\bf II} to the GL model, confined between two parallel planes and 
a study of the critical behaviour of the system is performed. In particular, we obtain 
an expression for the critical temperature as a function of of the spacing between the
planes (the film thickness). Finally, we summarize our results in Section {\bf IV} 
and a qualitative comparison with some experimental observations is done.

\section{The Gaussian effective potential for the Ginzburg-Landau model}

We begin by briefly presenting the study of the U(1) Scalar
Electrodynamics in the transverse unitarity gauge,
along the lines developed in ref.\cite{Camarda}. 
We start from the Hamiltonian density of
the GL model in Euclidean $d$-dimensional space written in the form \cite{Kleinert},
\begin{eqnarray}
{\cal H'}&=& \frac{1}{4}F_{\mu\nu}F^{\mu\nu}+\frac{1}{2}|
(\partial_{\mu} -ieA_{\mu})\Psi |^{2} \nonumber \\
&+& \frac{1}{2}m_{0}^{2}|\Psi|^{2} +\lambda (|\Psi |^{2})^{2},
\label{GL}
\end{eqnarray}
where $\Psi$ is a complex field, and $m_{0}$ is the bare mass.
The components of the transverse magnetic field, 
$F_{\mu\nu}=\partial_{\mu}A_{\nu}-\partial_{\nu}A_{\mu}$
($\mu,\nu=1,...,d$) are related to the $d$-dimensional potential vector by the well 
known equation, 

\begin{equation}
\frac{1}{4}F_{\mu\nu}F^{\mu\nu}=|{\bf \nabla}\times {\bf A}|^{2}.
\label{gau}
\end{equation}
In order to obtain only physical degrees of freedom, we can introduce two real fields 
instead of the complex field $\Psi$, assuming a transverse unitarity gauge.
We can define the field in terms of two real fields, as $\Psi=\phi
e^{i\gamma}$, together with the gauge transformation ${\bf A}\rightarrow
{\bf A}-1/e {\bf \nabla}\gamma$. The unitarity gauge makes 
the original transverse field to acquire a longitudinal component
${\bf A}_{L}$ proportional to ${\bf \nabla}\gamma$. Then the original 
functional integration over $\Psi$ and $\Psi^{*}$ in the generating
functional of correlation functions,  becomes an integration over
$\phi,\; {\bf A}_{T}$ and ${\bf A}_{L}$. The longitudinal component of 
the vector potential can be integrated out, leading to 
the generating functional (up to constant terms),

\begin{equation}
Z[j]=\int D\phi\;DA_{T}]exp \left[ -\int d^{d}x {\cal H} + \int
d^{d}x \;j\phi \right],
 \label{part}
\end{equation}
where the Hamiltonian is

\begin{eqnarray}
{\cal H}&=&\frac{1}{2}(\nabla\phi)^2 + \frac{1}{2}m_{0}^{2}\phi^{2}
+\lambda\phi^{4} \nonumber \\
&+& \frac{1}{2}e^{2}\phi^{2}A^{2}+\frac{1}{2}({\bf
\nabla}\times {\bf A})^{2} + \frac{1}{2\epsilon}({\bf \nabla}
\cdot {\bf A})^{2}. \label{Lagrangeana}
\end{eqnarray}
We have introduced above a gauge fixing term, the limit $\epsilon \rightarrow 0$
being taken later on after the calculations have been done. 
In Eq.(\ref{Lagrangeana}) and in what follows, unless explicitly stated, 
${\bf A}$ stands for the {\it transverse} gauge field.

The Gaussian effective potential can be obtained  
from Eq.(\ref{Lagrangeana}), performing a  
shift in the scalar field in the form $\phi = \tilde{\phi} + \varphi $,
which allows to write the Hamiltonian in the form 
\be
{\cal H}={\cal H}_{0} + {\cal
H}_{int},
\label{H1}
\ee
 with ${\cal H}_{0}$ being the free part and ${\cal
H}_{int}$ the interaction part, given respectively by

\begin{eqnarray}
{\cal H}_{0}&=& \left[ \frac{1}{2}(\nabla
\tilde{\phi})^2+\frac{1}{2} \Omega^{2}\tilde{\phi}^{2} \right] \nonumber \\
&+& \left[\frac{1}{2}({\bf \nabla}\times {\bf A})^{2} +\frac{1}{2}
\Delta^{2}A_{\mu}A^{\mu} +\frac{1}{2\epsilon}({\bf \nabla} \cdot
{\bf A})^{2}\right], \label{H0}
\end{eqnarray}
and

\begin{eqnarray}
{\cal H}_{int} &=& \sum_{n=0}^{4}
v_{n}\tilde{\phi}^{n}+\frac{1}{2}\left(
e^{2}\varphi^{2}-\Delta^{2} \right)A_{\mu}A^{\mu} \nonumber \\
&+& \frac{1}{2}e^{2}\tilde{\phi} A_{\mu}A^{\mu}\varphi+\frac{1}{2}
e^{2}A_{\mu}A^{\mu}\varphi^{2}, \label{Hint}
\end{eqnarray}
where

\begin{eqnarray}
v_{0}&=&\frac{1}{2}m_{0}^{2}\varphi^{2}+\lambda\varphi^{4},\label{v0} \\
v_{1}&=&m_{0}^{2}\varphi+4\lambda\varphi^{3}, \label{v1} \\
v_{2}&=&\frac{1}{2}m_{0}^{2}\varphi^{2}+6\lambda\varphi^{2}
-\frac{1}{2}\Omega^{2},  \label{v2} \\
v_{3}&=&4\lambda\varphi,  \label{v3} \\
v_{4}&=&\lambda. \label{v4}
\end{eqnarray}
It is clear from Eqs. (\ref{H1}, 
(\ref{H0}) and (\ref{Hint}), that ${\cal H}$ describes two interacting fields, a
real scalar field $\phi$ of mass $\Omega$ and a real vector
gauge field {\bf A} of mass $\Delta$.

The effective potential, which is defined by
\begin{equation}
V_{eff}[\varphi]= \frac{1}{V}\left[ -lnZ + \int d^{d}x j\varphi
\right], \label{defpot}
\end{equation}
where $V$ is the total volume, can be obtained at first 
order from standard methods from perturbation
theory. One can find, from Eqs. (\ref{part}),(\ref{H0}) and
(\ref{Hint}),

\begin{eqnarray}
V_{eff}[\varphi]&=& I_{1}^{d}(\Omega) + 2I_{1}^{d}(\Delta) +
\frac{1}{2}m_{0}^{2}\varphi^{2}+\lambda\varphi^{4} \nonumber\\
&& + \frac{1}{2} \left[ m_{0}^{2} - \Omega^{2}+
12\lambda\varphi^{2}+6\lambda I_{0}^{d}(\Omega) \right]
I_{0}^{d}(\Omega)\nonumber \\
&& + \left[ e^{2} \left( \varphi^{2} + I_{0}^{d}(\Omega)\right) -
\Delta^{2} \right]I_{0}^{d}(\Delta), \label{poteff}
\end{eqnarray}
where, 

\be
I_{0}^{d}(M) = \int \frac{d^{d}k}{(2\pi)^{d}}\frac{1}
{k^{2}+M^{2}}, 
\label{I0}
\ee
\be
I_{1}^{d}(M) = \frac{1}{2} \int \frac{d^{d}k}{(2\pi)^{d}}
\ln(k^{2}+M^{2}), 
\label{I1}
\ee
with $k=(k_{1},...,k_{d})$ being the $d$-dimensional momentum.

The Gaussian effective potential is derived by the requirement that $V_{eff}[\varphi]$
must be stationary under variations of the masses $\Delta $ and
$\Omega $. This means that values $\overline{\Omega}$ and $\overline{\Delta}$ 
for the masses $\Omega$ and $\Delta$ should be found such that the conditions,

\begin{eqnarray}
\left. \frac{\partial V_{eff}}{\partial
\Omega^{2}}\right|_{\Omega^{2}=\overline{\Omega}^{2}}
 &=& 0, \label{sc1}\\
\left. \frac{\partial V_{eff}}{\partial
\Delta^{2}}\right|_{\Delta^{2}= \overline{\Delta}^{2}}&=& 0,
\label{sc2}
\end{eqnarray}
 be simultaneously satisfied. 
  These conditions generate the gap equations,

\begin{eqnarray}
\overline{\Omega}&=& m_{0}^{2} + 12\lambda \varphi^{2} +12\lambda
I_{0}^{d}(\overline{\Omega}) + 2e^{2} I_{0}^{d}
(\overline{\Delta}), \label{gap1} \\
\overline{\Delta}&=& e^{2} \varphi^{2}+ e^{2}
I_{0}^{d}(\overline{\Omega}). \label{gap2}
\end{eqnarray}
Replacing  $\Omega$ and $\Delta$ in Eq.(\ref{poteff})
by the solutions $\overline{\Omega}$ and $\overline{\Delta}$, of Eqs.(\ref{gap1}) and 
(\ref{gap2}) we obtain for the GEP the formal expression,

\begin{eqnarray}
\overline{V}_{eff}[\varphi] &=& I_{1}^{d}(\overline{\Omega}) +
2I_{1}^{d}(\overline{\Delta})+\frac{1}{2} m_{0}^{2}\varphi^{2}
+\lambda\varphi^{4} \nonumber\\
&-& 3\lambda [I_{0}^{d}(\overline{\Omega})]^{2}-e^{2}
I_{0}^{d}(\overline{\Omega})I_{0}^{d}(\overline{\Delta}).
\label{GEP}
\end{eqnarray}
Notice that Eqs.(\ref{gap1}) and (\ref{gap2}) are a pair of very involved coupled
equations, and no analytical solution for 
them has been found, they can be solved only by numerical methods.
Later on we will see that this difficulty, in the limit of criticality can be
surmounted.

Next we intend to write an expression for the Gaussian
mass, $\overline{m}$, obtained in our case from the standard
prescription, as the second derivative of the {\it Gaussian} effective potential 
for $\varphi=0$. To calculate the second derivative
of $\overline{V}_{eff}$ with respect to $\varphi$, we remark from
Eqs.(\ref{gap1}) and (\ref{gap2}) that $\overline{\Omega}^{2}$ and
$\overline{\Delta}^{2}$ also depend on $\varphi$ according to the
relations

\begin{eqnarray}
\frac{d\overline{\Omega}^{2}}{d\varphi} &=& \frac{24\lambda\varphi
-e^{2}I_{-1}^{d}(\overline{\Delta})\frac{d\overline{\Delta}^{2}}
{d\varphi}}{1+6\lambda I_{-1}^{d}(\overline{\Omega})}, \label{rel1}\\
\frac{d\overline{\Delta}^{2}}{d\varphi} &=& 2e^{2}\varphi
-\frac{1}{2}e^{2}I_{-1}^{d}(\overline{\Delta})
\frac{d\overline{\Omega}^{2}}{d\varphi}, \label{rel2}
\end{eqnarray}
where

\begin{equation}
I_{-1}^{d}(M)=2 \int \frac{d^{d}k}{(2\pi)^{d}}
\frac{1}{(k^{2}+M^{2})^{2}}. \label{I-1}
\end{equation}
Replacing Eq.(\ref{rel2}) in (\ref{rel1}) we get, 

\begin{equation}
\frac{d\overline{\Omega}^{2}}{d\varphi} = \frac{\left[ 24\lambda
-2e^{4}I_{-1}^{d}(\overline{\Delta})\right] \varphi} {1+\left[
6\lambda - \frac{1}{2}e^{4}I_{-1}^{d}(\overline{\Omega}) \right]
I_{-1}^{d}(\overline{\Omega})}, \label{rel3}
\end{equation}
and the second derivative of the GEP with respect to $\varphi$ is given by, 
\begin{eqnarray}
\frac{d^{2} \overline{V}_{eff}}{d \varphi^{2}}&=& m_{0}^{2}+12\lambda
\varphi^{2} \nonumber \\
&+& 12\lambda I_{0}^{d}(\overline{\Omega})+2e^{2}I_{0}^{d}
(\overline{\Delta})+2 e^{4} \varphi^{2}I_{-1}^{d}(\overline{\Delta})
\nonumber \\
&-&\frac{\left[ 6\lambda + \frac{1}{2}
e^{4}I_{-1}^{d}(\overline{\Delta}) \right]\left[
24\lambda-2e^{4}I_{-1}^{d}(\overline{\Delta})\right]
\varphi^{2}}{1+\left[ 6\lambda - \frac{1}{2}e^{4}I_{-1}
(\overline{\Omega}) \right]I_{-1}^{d}(\overline{\Omega})}. \nonumber\\
\label{2derV}
\end{eqnarray}
Thus we have the formula for the Gaussian mass,

\begin{eqnarray}
\overline{m}^{2} & \equiv & \left. \frac{d^{2} V_{eff}}{d
\varphi^{2}}\right|_{\varphi=0}= \nonumber \\
&=& m_{0}^{2}+ 12\lambda I_{0}^{d}(\overline{\Omega}_{0})
+2e^{2}I_{0}^{d}(\overline{\Delta}_{0}),  \label{grm}
\end{eqnarray}
where $\overline{\Omega}_{0}$ and $\overline{\Delta}_{0}$ are respectively solutions
for $\overline{\Omega}$ and $\overline{\Delta}$ at $\varphi=0$, explicitly, 

\begin{eqnarray}
\overline{\Omega}_{0}^{2}&=& m_{0}^{2} +12\lambda I_{0}^{d}
(\overline{\Omega}_{0})
+ 2e^{2} I_{0}^{d}(\overline{\Delta}_{0}),   \label{gap3} \\
\overline{\Delta_{0}}&=& e^{2}I_{0}^{d}(\overline{\Omega}_{0}).
\label{gap4}
\end{eqnarray}
Therefore, from Eqs.(\ref{grm}) and (\ref{gap3}) we get simply,
\begin{equation}
\overline{m}^{2}=\overline{\Omega}_{0}^{2}.  \label{mOm}
\end{equation}
Hence, we see from the gap equation
(\ref{grm}) that the Gaussian  mass obeys a generalized "Gaussian" 
Dyson-Schwinger equation,  
\begin{equation}
\overline{m}^{2}=m_{0}^{2}+ 12\lambda I_{0}^{d}(\overline{m})
+2e^{2}I_{0}^{d}\left(\sqrt{e^{2}I_{0}^{d}(\overline{m})}\right).
\label{gap5}
\end{equation}
This expression will be used later to describe the system in the neighbourhood of 
criticality.

\section{Critical behaviour of the confined Ginzburg-Landau model}

\subsection{The effect of confinement}

Let us now consider
the system confined between two parallel planes, normal to the $x_{d}$-axis, a
distance $L$ apart from one another and use Cartesian coordinates ${\bf r} 
=(x_{d},{\bf z})$, where ${\bf z}$ is a $(d-1)$-dimensional vector, with
corresponding momenta ${\bf k}=(k_{d},{\bf q})$, ${\bf q}$ being a $(d-1)$ 
-dimensional vector in momenta space. In this case, the model is supposed 
to describe a superconducting material in the form of a film.
Under these conditions the field $\phi(x_{d},{\bf z})$ satisfies the 
condition of confinement along
the $x_{d}$-axis, $\varphi(x_{d}=0, {\bf z})\;=\;\varphi(x_{d}=L, {\bf z})\;=\;0$, 
and should  
 have a mixed series-integral Fourier expansion of the 
form,
\begin{equation}
\phi(x_{d},{\bf z})=\sum_{n=-\infty}^{\infty} c_{n}\int d^{d-1} 
{\bf q} \;b({\bf q}) e^{-i\omega_{n} x_{d}\;-i{\bf q}\cdot {\bf z}}
\tilde{\varphi}(\omega_{n}, {\bf q}), 
\label{Fourier}
\end{equation}
where $\omega_{n}=2\pi n/L$ and the coefficients $c_{n}$ and $b({\bf q})$   
correspond respectively to the Fourier series representation over $x_{d}$ and 
  to the Fourier integral representation over the $(d-1)$-dimensional 
${\bf z}$-space.
The above conditions of confinement of the $x_{d}$-dependence of the
field to a segment of length $L$, allow us to 
proceed with respect to the $x_{d}$-coordinate, 
in a manner analogous as it is done in the imaginary-time 
Matsubara formalism in field theory. The Feynman rules should be modified 
following the prescription,
\begin{equation}
\int \frac{dk_{d}}{2\pi }\rightarrow \frac{1}{L}\sum_{n=-\infty }^{+\infty
}\;,\;\;\;\;\;\;k_{d}\rightarrow \frac{2n\pi }{L}\equiv \omega _{n}.
\label{Matsubara}
\end{equation}
We emphasize however, that here we are considering an
Euclidean field theory in $d$ {\it purely} spatial dimensions, we are {\it not} 
working in the framework of finite temperature field theory.   
Temperature is introduced in the mass term of the Hamiltonian by means of the 
usual Ginzburg-Landau prescription.

For our purposes we only need the calculation of the
integral given in Eq.(\ref{I0}) in the situation of confinement of the present section. 
With the prescription (\ref{Matsubara}), the equation corresponding to Eq.(\ref{I0}) 
for the confined system can be written in the form,
\begin{equation}
I_{0}^{d}(M) = \frac{1}{4\pi^{2}L}\sum_{n=-\infty }^{+\infty }\int
\frac{d^{d-1}q} {q^{2}+an^{2}+c^{2}}, \label{I00}
\end{equation}
where $q_{i}=k_{i}/2\pi$, $a=1/L^{2}$ and $c^{2}=M^{2}/4\pi^{2}$.

Eq.(\ref{I00}) can be treated within the framework of the formalism developed in
Refs. \cite{Ademir},\cite{JMario}. Using a well known regularization
formula \cite{Ramond}, we can write Eq.(\ref{I00}) in the form
\begin{equation}
I_{0}^{d}(M) = \frac{\sqrt{a}}{4\pi^{2-d/2}}\Gamma\left(
2-\frac{d}{2} \right) A_{1}^{c^{2}}\left(\frac{3-d}{2},a\right),
\label{I01}
\end{equation}
where $A_{1}^{c^{2}}\left(\frac{3-d}{2},a\right)$ is one of the
Epstein-Hurwitz zeta-functions, defined by\cite{Elizalde} 
\begin{equation}
A_{K}^{c^{2}}(\nu;\{a_{i}\})=\sum_{n_{1},...,n_{K}=-\infty
}^{+\infty }(a_{1}n_{1}^{2}+...+a_{K}n_{K}^{2}+c^{2})^{-\nu},
\label{zeta1}
\end{equation}
with $Re(\nu)>\;K/2$ (in our case $Re(d)<\;2$). The
Epstein-Hurwitz $zeta$-function can be extended as a meromorphic function to the whole
complex $\nu$-plane (for us, to all values of the 
dimension $d$), and we obtain after some rather long but 
straightforward manipulations described in detail in \cite{Ademir}, the
expression,
\begin{eqnarray}
I_{0}^{d}(M) &=& 2^{-\frac{d}{2}}\pi^{1-\frac{d}{2}}\left[
2^{1-\frac{d}{2}}\Gamma\left( 1-\frac{d}{2} \right) M^{-2+d}
 \right. \nonumber\\
&& +\left. 2\sum_{n=1}^{\infty}\left( \frac{M}{nL} \right)^{-1+
\frac{d}{2}}K_{-1+ \frac{d}{2}}(MLn) \right], 
\label{I02}
\end{eqnarray}
where $K_{\nu}$ are the Bessel functions of third kind.

\subsection{Critical behaviour}

Let us take $M=\overline{m}$ in Eq.(\ref{I02}) and  
 let us restrict ourselves to the neighbourhood of criticality, that is, to the region 
defined by $\overline{m} \approx 0$. Then the asymptotic formula,
\begin{equation}
K_{\nu }(z)\approx \frac{1}{2}\Gamma (\nu )\left(
\frac{z}{2}\right) ^{-\nu },\;\;\;(z\sim 0)  \label{K}
\end{equation}
allows to write Eq.(\ref{I02}) in the form
\begin{equation}
I_{0}^{d}(\overline{m} \approx 0)\approx \frac{\pi^{1-\frac{d}{2}}}{2} \Gamma\left(
1-\frac{d}{2} \right)\frac{1}{L^{d-2}}\zeta(d-2), \label{I03}
\end{equation}
where $\zeta (d-2)$ is the Riemann $zeta$-function, $\zeta
(d-2)=\sum_{n=1}^{\infty }(1/n^{d-2})$, defined for $d>3$.
For
$d \gtrsim 3$, in the sense of the analytic continuation in dimension 
of the Espstein-Hurwitz $zeta$-functions, we obtain the  expression, 
\begin{equation}
I_{0}^{d}(\overline{m}\approx0) \approx  \frac{1}{2\sqrt{\pi}}
\frac{1}{L}\zeta(d-2). \label{I04}
\end{equation}

The integral $I_{0}^{d}\left(\overline{\Delta}_{0}=
\sqrt{e^{2}I_{0}^{d} (\overline{m})}\right)$, which enters 
Eq.(\ref{gap5}), must be considered carefully. For a dimension $d\gtrsim 3$,
we get, 
\begin{eqnarray}
I_{0}^{d}(\overline{\Delta}_{0}) &\approx & 2^{-\frac{3}{2}}
\pi^{-\frac{1}{2}}\left[ 2^{-\frac{1}{2}}\Gamma\left( -\frac{1}{2}
\right) \overline{\Delta}_{0}
 \right. \nonumber\\
&& +\left. 2\sum_{n=1}^{\infty}\left( \frac{\overline{\Delta}_{0}}
{nL} \right)^{ \frac{1}{2}}K_{\frac{1}{2}}
(\overline{\Delta}_{0}Ln) \right], \label{I05}
\end{eqnarray}
or, using the exact expression for the summation in the above equation,
\begin{equation}
\sum_{n=1}^{\infty}\left( \frac{\overline{\Delta}_{0}} {nL}
\right)^{ \frac{1}{2}}K_{\frac{1}{2}}(\overline{\Delta}_{0}Ln)=
-\sqrt{\frac{\pi}{2}}\frac{1}{L}\ln{\left(1-e^
{-\overline{\Delta}_{0}L}\right)}, 
\label{fK}
\end{equation}
Eq.(\ref{I05}) becomes (with $\overline{\Delta}_{0}=
\sqrt{e^{2}I_{0}^{d} (\overline{m})}$)
\begin{eqnarray}
I_{0}^{d}\left(\sqrt{e^{2}I_{0}^{d} (\overline{m})}\right) &\approx &
\frac{1}{2\sqrt{\pi}} \left[ \frac{1}{\sqrt{2}}\Gamma\left(
-\frac{1}{2} \right) \sqrt{e^{2}I_{d}^{d} (\overline{m})}
 \right. \nonumber\\
&&
\left.-\sqrt{2\pi}\frac{1}{L}\ln{\left(1-e^{-\sqrt{e^{2}I_{0}^{d}
(\overline{m})}L}\right)}\right]\nonumber \\
 \label{I06}
\end{eqnarray}
However, notice that if we are in the limit $\overline{m}\approx0$, we see 
replacing $I_{0}^{d}
(\overline{m})$ 
from Eq.(\ref{I04}) in the exponential, that the logarithm in the last term of
Eq.(\ref{I06}) will disappear at $d=3$, due to the divergence of $\zeta(d-2)$ as 
$d\rightarrow 3$. Hence, Eq.(\ref{I06}) 
becomes simply, for a dimension $d \gtrsim 3$,
\begin{equation}
I_{0}^{d}\left( \sqrt{e^{2}I_{0}^{d} (\overline{m}\approx0)} \right)
\approx \frac{e}{2\pi^{1/4}\sqrt{2}} \frac{1}{L^{\frac{1}{2}}}
\zeta^{\frac{1}{2}}(d-2). \label{I07}
\end{equation}
Thus we can write the Gaussian gap equation (\ref{gap5}) in the neighbourhood of 
criticality in the form,
\begin{equation}
\overline{m}^{2}\approx m_{0}^{2}+ \frac{24}{\sqrt{\pi}}\lambda \frac{1}{L}\zeta(d-2)
-\frac{1}{\pi^{1/4}\sqrt{2}}e^{3}\frac{1}{L^{\frac{1}{2}}}
\zeta^{\frac{1}{2}}(d-2). \label{crit1}
\end{equation}
For $\overline{m}=0$, Eq.(\ref{crit1}) defines a critical equation for $d\gtrsim 3$. 
But it is well known that the {\it only} 
singularity of the $zeta$-function $\zeta(z)$ is a pole at $z=1$, 
 which would make Eq.(\ref{crit1}) 
meaningless as it stands for $d=3$, 
just the physically interesting situation. 

However, we can give a physical sense
to  Eq.(\ref{crit1}) for $d=3$, by means of a 
 regularization procedure. This can be done using the 
formula, 
\begin{equation}
\lim_{z\rightarrow 1}\left[ \zeta (z)-\frac{1}{1-z}
\right]=\gamma, \label{zeta}
\end{equation}
where $\gamma$ is the Euler constant,
to define for $d\gtrsim 3$ a new, {\it dressed} mass $m$, related to 
the former bare mass by,
\ben
m^{2}&=& m_{0}^{2}-\frac{24\lambda}{\sqrt{\pi}L(d-3)}+\nonumber \\ & & +\frac{e^{3}}{2\pi^{1/4}\sqrt{2L}}\sum_{p=1}^{\infty}C_{\frac{1}{2}}^{p}
\gamma^{\frac{1}{2}-p}\frac{(-1)^{p}}{(d-3)^{p}},
\label{mr}
\een
where $C_{\frac{1}{2}}^{p}$ are appropriate generalizations of the 
 coefficients of the binomial expansion for a fractional power. Then replacing
the above equation in Eq.(\ref{crit1}) and using
the binomial formula to expand $\zeta^{1/2}(d-2)\approx \left[\gamma 
-(1/(d-3)\right]^{1/2}$ we obtain, for $d=3$, 
\be
\overline{m}^{2}\approx m^{2}+\frac{24}{\sqrt{\pi}}\lambda
\frac{1}{L}\gamma
-\frac{1}{\pi^{1/4}\sqrt{2}}e^{3}\frac{\gamma^{\frac{1}{2}}}{L^{\frac{1}{2}}}.
\label{zeta4}
\ee
Taking $m^{2}=\alpha (T-T_{0})$, 
with $\alpha >0$ we have from the above equation for $\overline{m}^{2}=0$, 
the critical temperature as a function of the film thickness $L$ and of the bulk 
transition temperature, $T_{0}$,
\begin{equation}
T_{c}=T_{0}- \frac{24\beta \gamma}{\alpha \sqrt{\pi}}
\frac{1}{L}
+\frac{e^{3}\sqrt{\gamma}}{\alpha \pi^{1/4}\sqrt{2}}\frac{1}{\sqrt{L}}. 
\label{crit4}
\end{equation}
where, in agreement with the standard notation, we have introduced 
  $\beta=\lambda$, the Ginzburg-Landau parameter. This is an 
equation that describes the behaviour of the critical temperature
of a superconducting film of thickness $L$, taking into account the
gauge fluctuations. Of course when $L\rightarrow\infty$, we
recover the case of the material in bulk form.

 We see clearly two separated contributions in the expression 
to the critical temperature given in Eq.(\ref{crit4}). The first one due to the 
self-interaction of the scalar field, and the other coming from the interaction between 
the scalar and gauge fields. This last one would characterize a {\it charged} phase 
transition. It should be noticed that the self interaction contribution to the critical 
temperature 
depends on the inverse of the film thickness, while the charged contribution goes 
with the inverse of the square root of $L$. 
Taking $T_{c}=0$ in Eq.(\ref{crit4}), we obtain a positive solution for $L$, 
\begin{eqnarray}
L^{(0)}&=&\left(\frac{48\beta\gamma}{\alpha}\right) \left[
\frac{\sqrt{\gamma}}{\alpha\sqrt{2}\pi^{1/4}}e^{3}+ \left(
\frac{\gamma}{\alpha^(2)2\sqrt{\pi}}e^{6} \right.\right.
\nonumber \\ && \left.\left. +\frac{96\beta\gamma}
{\alpha \sqrt{2}\pi^{3/4}}T_{0}\right) ^{\frac{1}{2}}
\right]^{-2}.  
\label{lmin}
\end{eqnarray}
For $L<L^{(0)}$, the critical temperature (in absolute units) becomes negative, 
meaning that $L^{(0)}$ is the minimal physically allowed film thickness, below which the
superconducting transition is suppressed.

\section{Conclusions}
In this paper we have considered the Ginzburg-Landau model, confined 
between two parallel planes, and in the transverse unitarity gauge, as a model to describe a
superconducting film. To generate the contributions from gauge
fluctuations, we have used the Gaussian effective potential, which allows to obtain
a gap equation that can be treated with the method of recent developments
\cite{Ademir,JMario}. We have derived a critical equation that
describes the  changes in the critical temperature due to
 confinement. Independent contributions from the
self interaction of the scalar field and from the gauge field
fluctuations are found. Our approach suggests a minimal film thickness for
the existence of both charged and non-charged superconducting transitions.
The behaviour described in Eqs.(\ref{crit4}) 
and (\ref{lmin}) may be contrasted with the linear decreasing
of $T_{c}$ with the inverse of the film thickness, 
that has been found {\it experimentally} in materials
containing transition metals, for example, in Nb \cite{Nb3}, in
W-Re alloys \cite{W-Re} and in epitaxial MgB$_{2}$ films\cite{Xi};
for some of these cases, this behaviour has been explained in
terms of proximity, localization and Coulomb-interaction effects. With 
our formalism such a linear 
decreasing of $T_{c}$ with the inverse film thickness, can be directly obtained 
from our Eq.(\ref{crit4}), simply taking $e=0$ (this could mean that the transitions 
observed in the above quoted references are non-charged). Also a comparison may be done with recent 
theoretical results for type II superconductors\cite{Luciano}, where 
a similar behaviour of the critical temperature with the film thickness 
has been found for non-charged transitions. 
Moreover we would like to emphasize, that our results do not depend on microscopic
details of the material involved nor account for the influence of
manufacturing aspects, like the kind of substrate on which the
film is deposited. In other words, our results emerge solely as a property that
appears in the context of 
the Gaussian effective potential formalism and as a 
topological effect of the compactification of the Ginzburg-Landau
model in one direction. 
Finnaly we would like to remark that our approach includes both 
{\it charged} and non-charged transitions.  
 Our results are in qualitative agreement with 
 the experimentally observed behaviours mentioned above, 
under the assumption that they correspond to non-charged phase transitions.\\

{\bf Acknowledgements}

This work has been supported by CNPq (Brazilian national research council); one of us 
(I.R.) also thanks PRONEX.

\end{document}